\documentclass[fleqn,10pt]{wlscirep}
\usepackage[utf8]{inputenc}
\usepackage[T1]{fontenc}
\usepackage{xcolor}
\usepackage{comment}
\usepackage{lastpage}

\usepackage{here} 
\newenvironment{figurehere}
  {\begin{figure}[H]}
  {\end{figure}}

\usepackage[percent]{overpic} 

\usepackage{xr}

\title{Do Researchers Benefit Career-wise from Involvement in International Policy Guideline Development?}

\author[1,*]{Yuta Tomokiyo}
\author[1]{Keita Nishimoto}
\author[1]{Kimitaka Asatani}
\author[1]{Ichiro Sakata}

\affil[1]{Department of Technology Management for Innovation, Graduate School of Engineering, The University of Tokyo, Tokyo 113-8656, Japan}
\affil[*]{Corresponding author, y.tomo.research@gmail.com, tomokiyo-yuta721@g.ecc.u-tokyo.ac.jp}

\begin{abstract}
Researchers are no longer limited to producing knowledge; in today's complex world, they also address societal challenges by engaging in policymaking. Although involvement in policymaking has expanded, direct empirical evidence of its career benefits remains underexplored. Prior survey-based studies suggest potential advantages--such as broader professional networks and enhanced opportunities--yet raise concerns about insufficient institutional support. Here, we examine the 2021 WHO global air quality guideline--a science-based regulatory guideline--as a case study. To evaluate the impact of guideline development on research outcomes, we match guideline researchers with a control group of peers sharing similar research topics and prior performance. Our analysis reveals that guideline researchers attain higher future citation counts in both academic and policy domains. New collaborations formed during development yield publications with higher citation impact and the disruptive index. Moreover, about half the guideline's references are derived from guideline researchers' papers, highlighting their central role in shaping the evidence base. These results provide empirical support for the career benefits of policy engagement. Our findings indicate that engaging in international guideline development offers tangible career incentives for researchers, and that institutions can enhance research impact and promote innovative scientific progress by actively supporting their researchers' participation in such initiatives.
\end{abstract}

\begin{document}
\flushbottom
\maketitle

\thispagestyle{empty}

\section*{Introduction}

Researchers are no longer limited to producing knowledge; in today's increasingly complex world, they are also called upon to provide evidence for pressing social issues and, in many cases, to engage directly in policymaking \cite{lubchenco1998entering, guston2001boundary, lach2003advocacy, scott2007policy}. Tackling global complex problems such as climate change \cite{demeritt2001construction, van2022navigating}, emerging infectious diseases\cite{yin2021coevolution, bornmann2022relevant}, and air pollution \cite{vilcassim_GapsFutureDirections_2023b} requires more than just accumulating scientific insight--it demands effectively translating that insight into policy decisions\cite{guston2001boundary, jagannathan2023research}. COVID-19 pandemic, for example, made it clear that evidence-based policymaking can be decisive for public health measures and that researchers' expertise can directly impact decision-making \cite{hodges2022role}. As a result, there is a growing recognition of researchers' dual role as both knowledge producers and pivotal intermediaries between science and society \cite{jagannathan2023research}.

Despite this heightened awareness of researchers' policy engagement, empirical evidence on how it affects their careers remains limited. Survey-based studies suggest that researchers involved in policymaking often perceive potential benefits such as broader professional networks, expanded career opportunities, and skill development \cite{singh2019researcher, filyushkina2022engaging}. However, many still express concern that these activities receive insufficient credit or tangible rewards\cite{singh2019researcher}, additionally they are in the face of already demanding workloads\cite{lashuel2020busy}. Clarifying whether there are tangible career advantages--and, if so, identifying them--could provide a key incentive for broader involvement in policy engagement. However, much of the available research has relied on questionnaire or perception-based data, leaving a gap in empirically grounded insights into how policy engagement shapes researchers' academic careers \cite{singh2019researcher, filyushkina2022engaging}.

In this study, we focus on the process of developing international policy guidelines--undertaken by intergovernmental organizations (IGOs) such as the United Nations (UN) and the World Health Organization (WHO)--as vital channels through which scientists translate evidence into policymaking, and we examine how participation in these processes affects researchers' careers, particularly their impact on the scientific community and policymaking during and after guideline involvement. According to the Overton database, policy guidelines are one type of policy paper--others include government documents, IGO/NGO reports, think tank research, central bank working papers and more\cite{overton_index}. Developing these guidelines typically involves teams of experts who review and synthesize the most up-to-date research. By contributing to guideline development, researchers may not only strengthen the scientific basis of policy decisions but also cultivate new networks and potentially boost their academic influence. Previous empirical studies have highlighted selection biases among scholars contributing to Intergovernmental Panel on Climate Change (IPCC) reports \cite{corbera2016patterns} and have shown that research cited in climate policy documents tends to garner numerous academic citations \cite{bornmann2022relevant}. Nonetheless, direct evidence connecting guideline involvement to researchers' subsequent research performance remains limited.

To address this gap, we leverage Scopus--a comprehensive academic information database--and Overton--a database that includes policy documents and their citation information--to analyze whether researchers who involved in an international policy guideline development experienced changes in their academic trajectories. First, as Fig\ref{fig_yearly_citation} shows, we pair these researchers with a control group of peers in the same field and with comparable performance backgrounds, assessing whether guideline involvement correlates with growth in academic and policy citations. Second, we investigate whether the guideline fosters new collaborations and, if so, whether papers arising from these collaborations achieve higher citation counts and greater innovativeness. Finally, we assess the extent to which the researchers' own publications are cited within the guideline, thereby evaluating their direct influence on policy. As a case study, we focus on the WHO global air quality guideline--a science-based regulatory guideline--issued in 2021 \cite{world2021global}, which was developed from 2016 to 2021. By tracking changes in academic and policy citations--as well as shifts in collaborative networks--over time, we provide quantitative, causally informed insights into the career benefits that may arise from policy engagement. Our findings indicate that researchers involved in the guideline process experienced a significant increase in both academic and policy citations in the future and that new research collaborations established during this period yielded disruptive and high citation papers. Additionally, about half proportion of the citations within the guideline also originated from the researchers' own publications, underscoring their central role in shaping the evidence base for policymaking. 

\section*{The Guideline Development Process and Selection of Researchers for Analysis}

In this study, we focus on the international air quality guideline published by WHO in 2021\cite{world2021global}. Air quality and atmospheric pollution are critical public health issues, with air pollution linked to an estimated 6.7 million premature deaths worldwide\cite{fuller2022pollution}. This guideline was developed over a six-year period from 2016 to 2021. The guideline development involved 131 individuals assigned to specific expert roles--namely, WHO Steering Group, Guideline Development Group, Systematic Review Team, External Methodologists, and External Review Group. Guideline researchers are made up of a diverse range of nationalities, with a high proportion from Europe and the United States (Fig.~\ref{fig:guideline_basic_info}a). For our analysis, we define \textit{guideline researchers} as the 111 experts who were directly involved in the creation of the guideline's content (excluding the WHO Steering Group). Among these 111 experts, 108 were successfully matched to the Scopus dataset (see Methods section for details).

\section*{Assessing the Effect of Guideline Development on Academic and Policy Citations}

\begin{figure}[t]
    \centering
    \includegraphics[width=1.0\linewidth]{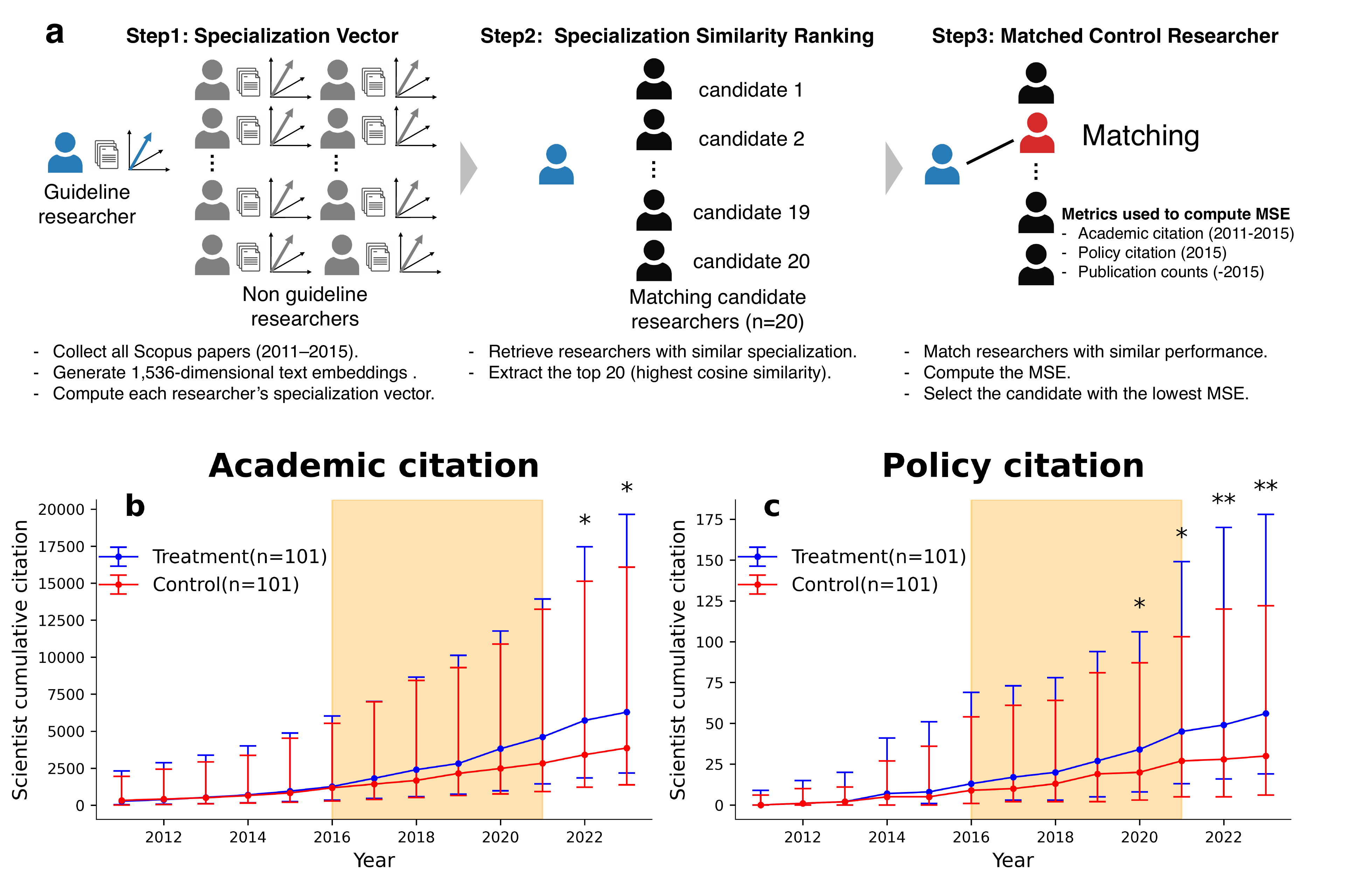}
    \caption{\textbf{a} Flow diagram of the matching procedure used to match guideline researchers with control researchers. In Step 1, each guideline researcher's \textit{specialization vector} is computed by averaging 1536-dimensional text embeddings derived from titles or abstracts of the papers authored by them in Scopus (2011--2015). In Step 2, the top 20 non-guideline researchers with the highest cosine similarity to each specialization vector are retrieved. In Step 3, the mean squared error (MSE) is calculated based on academic citations (2011--2015), IGO policy citations (as of 2015), and publication counts (as of 2015), and the candidate with the lowest MSE is selected as the matched control researcher. \textbf{b--c} Trends in cumulative academic and IGO policy citations for the treatment and control groups from 2011 to 2021. Orange marks the guideline development period (2016--2021). The plots show the median values for each group per year, with error bars indicating interquartile range. ** $p<0.01$, * $p<0.05$. }
    \label{fig_yearly_citation}
\end{figure}

To evaluate the impact of guideline development on research outcomes, we compared two groups (Fig.~\ref{fig_yearly_citation})--control researchers (n=101) and guideline researchers (n=101, treatment group)--who were matched based on research topics, cumulative academic citation counts (2011--2015), cumulative IGO policy citation counts (as of 2015), and publication numbers (as of 2015) (see Methods section for details). Of the 108 guideline researchers, 7 could not be matched because they had not published any papers between 2011 and 2015 or had published only in journals that were not indexed by Scopus. As illustrated in Fig.~\ref{fig:matching_metrics}, the distributions of the matching variables--academic citations, policy citations, and publication counts (Step 3)--did not differ significantly between the treatment and control groups (see also Supplementary), confirming that our control group was well-chosen.

We then examined whether differences in citation performance emerged over time. Specifically, we tracked each researcher's cumulative academic and IGO policy citations on a yearly basis. Each indicator was log10-transformed to approximate normality \cite{radicchi2008universality}. However, although the transformed data still deviated from normality (Shapiro-Wilk test, \textit{p} < 0.05), we performed one-sided Mann-Whitney \textit{U} tests at a 5\% significance level. As shown in Fig~\ref{fig_yearly_citation}b--c, no significant differences in academic or IGO policy citations were observed between the two groups in 2015. From 2016 onward, however, the treatment group's citation performance increased considerably, with statistically significant differences evident in policy citations from 2020 and in academic citations from 2022. These results suggest that involvement in guideline development process enhance both academic and policy impact. 

\section*{Emerging New Collaborations among Guideline Researchers through Guideline Development Period}

\begin{figure}
    \centering
    \includegraphics[width=1.0\linewidth]{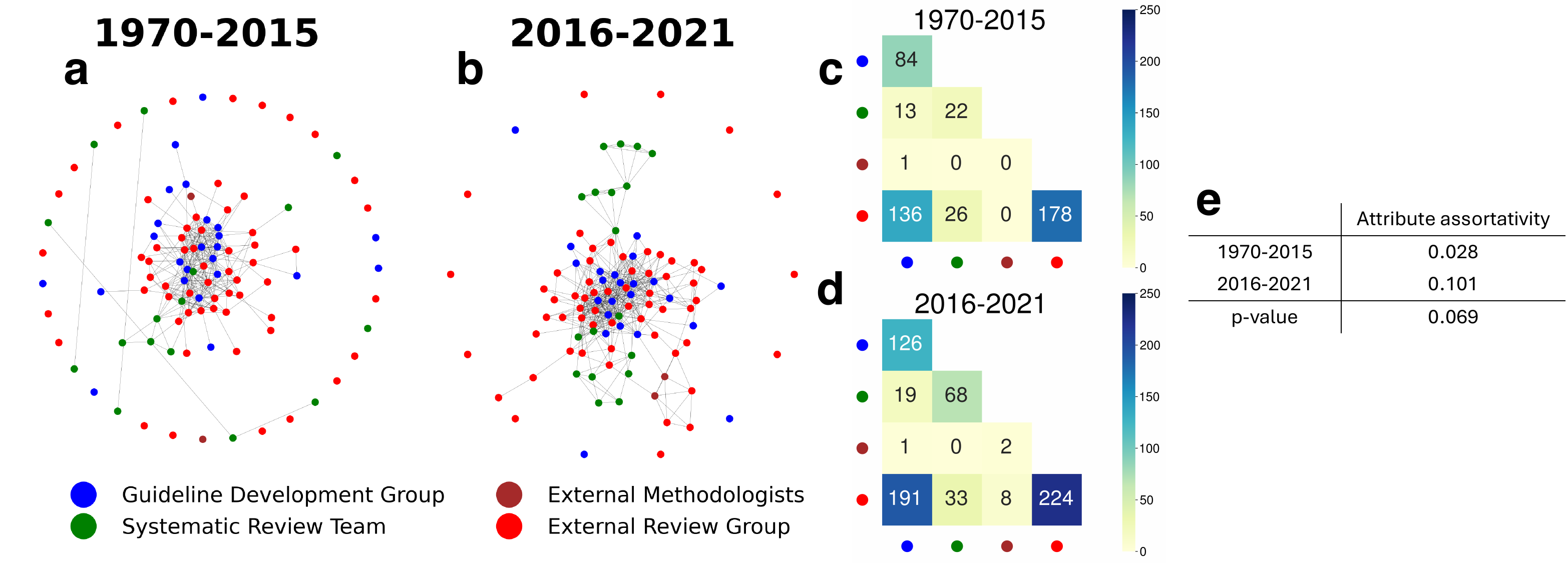}
   \caption{Co-authorship networks and attribute assortativity of guideline researchers. Node colors indicate roles: blue for the Guideline Development Group, green for the Systematic Review Team, brown for External Methodologists, and red for the External Review Group. \textbf{a} The co-authorship network constructed from publications spanning 1970--2015 (n=102). \textbf{b} The co-authorship network during the guideline development period (2016--2021) (n=104). \textbf{c} Total inter-attribute edge counts in the 1970--2015 network. \textbf{d} Total inter-attribute edge counts in the 2016--2021 network. \textbf{e} Comparison of attribute assortativity between the two periods networks, showing an increase from 0.028 (1970--2015) to 0.101 (2016--2021) with a borderline significant difference (\textit{p} = 0.069). }
   \label{fig:co-authorshipN}
\end{figure}

We next examined whether new collaborations formed among guideline researchers during the guideline development period. Fig~\ref{fig:co-authorshipN}a,b illustrates the co-authorship network of guideline researchers, comparing the period from 1970--2015 (panel a) with the guideline development period of 2016--2021 (panel b). To provide a baseline for this comparison, nodes in these networks are colored by the roles that researchers would eventually assume in the guideline development process. Notably, the largest connected component grew from 67 members (1970--2015) to 86 members (2016--2021), indicating a substantial increase in network connectivity that coincided with the guideline's creation.

In addition, a closer look at the two networks reveals the emergence of new co-authorship ties within the same roles (Fig.~\ref{fig:co-authorshipN}c,d)--for instance, within the Systematic Review Team (in green, Fig.~\ref{fig:co-authorshipN}b). Fig~\ref{fig:co-authorshipN}e summarizes the change in attribute assortativity\cite{newman2003mixing} (i.e., how likely researchers are to co-author within their same role), which rose from 0.028 (1970--2015) to 0.101 (2016--2021). Although this increase did not reach conventional significance (permutation test, \textit{p} = 0.069; see Methods section), the observed trend suggests that collaborative efforts across guideline teams have fostered additional academic partnerships.

\begin{figure}
    \centering
    \includegraphics[width=1.0\linewidth]{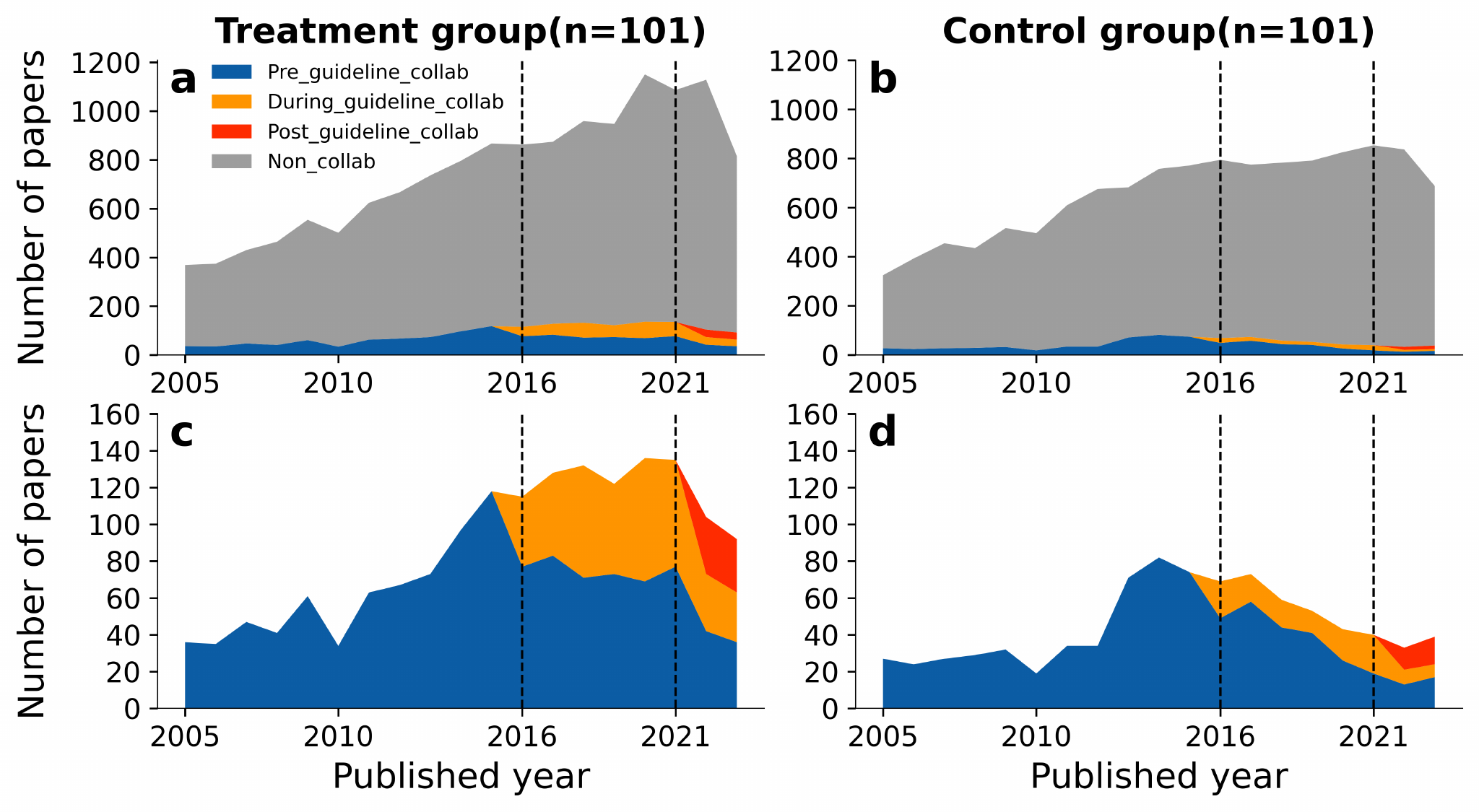}
    \caption{Trends in collaboration types among guideline researchers, treatment, and control groups.
    \textbf{a}--\textbf{b} show the total number of papers published each year (stacked by collaboration-initiation period), while \textbf{c}--\textbf{d} include only those papers co-authored within each respective group. \textbf{a}, \textbf{c} represent the treatment group (\(n=101\)), and \textbf{b}, \textbf{d} the control group (\(n=101\)). Vertical dashed lines mark the official guideline development window (2016--2021). Although the overall publication output increased for all groups, the treatment group (c) show a marked surge in new intra-group collaborations (orange) after 2016, in contrast to the control group (d).}
    \label{fig:collab}
\end{figure}

To investigate whether new collaborations emerged during the guideline development period, we focused on the timing of the first collaboration among researchers within each group, comparing the treatment group (n=101; Fig. \ref{fig:collab}a, c) and the matched control group (n=101; Fig. \ref{fig:collab}b, d) (see Methods section for details). The papers written by the researchers in each group were classified as non-collaborative (papers authored by the researcher alone or co-authored with researchers outside the group) or categorized based on the period when the co-author pair within the group first collaborated: "\textit{Pre\_guideline\_collab,} ," "\textit{During\_guideline\_collab,}," or "\textit{Post\_guideline\_collab}" (see Methods section).

Figs.~\ref{fig:collab}a--b show a clear expansion in total publications over time for both groups, but the composition of collaborations differs notably. The treatment group (Fig.~\ref{fig:collab}c) showed a pronounced rise in newly formed collaborations (highlighted in orange) during 2016--2021--particularly after the guideline process began. By contrast, the control group (Fig.~\ref{fig:collab}d) did not exhibit a similar upsurge. Although both groups had comparable levels of new collaborations prior to 2015, only those directly involved in the guideline development displayed a clear and expansion afterward. This finding suggests that involvement in guideline development process act as a catalyst for new academic partnerships, reshaping the collaborative landscape for involved researchers.

\section*{Exploring the Link between Collaboration Types and Research Impact}

We further explored whether different collaboration types affected the impact of papers published by guideline researchers (n=108) between 2016 and 2021. We evaluated two metrics of impact: the two-year citation count (\(C_2\)), log-transformed as \(\log_{10}(C_2 + 1)\) to approximate normality \cite{radicchi2008universality}, and the disruptive index (\(D\)), which captures the innovativeness of a publication \cite{funk_DynamicNetworkMeasure_2017, wu_LargeTeamsDevelop_2019}. To account for multiple comparisons, since some of the combinations of comparisons did not follow a normal distribution and homogeneity of variance, we applied one-sided Brunner-Munzel tests with Bonferroni correction, setting the significance threshold at \(p<0.0167\). 

\begin{figure}
    \centering
    \includegraphics[width=1.0\linewidth]{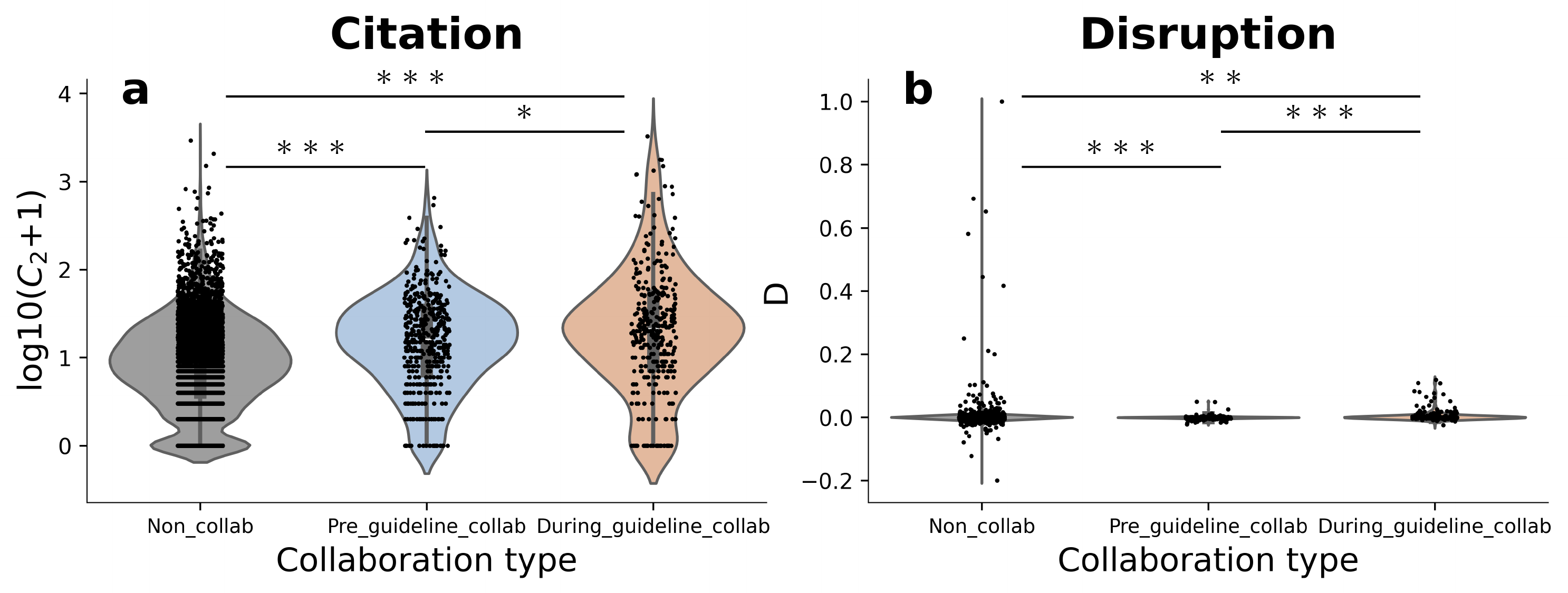}
    \caption{Impact of collaboration types on citation and disruption for papers published between 2016 and 2021 (the guideline development period).
\textbf{a}~Log-transformed two-year citation counts \(\log_{10}(C_{2}+1)\) for papers in each collaboration type (\textit{Non\_collab}, \textit{Pre\_guideline\_collab}, \textit{During\_guideline\_collab}). 
\textbf{b}~Disruptive index \((D)\) for the same collaboration types. 
Each violin plot shows the distribution for all papers published between 2016 and 2021 in that group, with individual points representing single publications. *** \(p<0.001\), ** \(p<0.01\), * \(p<0.0167\).}
    \label{fig:paper_index}
\end{figure}

Fig.~\ref{fig:paper_index}a compares the log-transformed citation counts across three collaboration types. The log-transformed citation distributions for all types deviated significantly from normality (Shapiro-Wilk test, \(p<0.05\)), and the variances were not homogeneous across all pairwise comparisons (Levene test, \(p<0.05\)). We observed that papers in the During\_guideline\_collab group (n=343) had significantly higher citation counts than those in both the Pre\_guideline\_collab group (n=447; one-sided Brunner-Munzel test, \(p<0.0167\), Cliff's Delta = 0.10) and the Non\_collab group (n=5,104; one-sided Brunner-Munzel test, \(p<0.001\), Cliff's Delta = 0.36). Furthermore, the Pre\_guideline\_collab group had significantly higher citation counts than the Non\_collab group (one-sided Brunner-Munzel test, \(p<0.001\), Cliff's Delta = 0.29). 

We then turned to the disruptive index (\(D\)), which also showed deviations from normality (Shapiro-Wilk test, \(p<0.05\)), and the variances were not homogeneous across all pairwise comparisons (Levene test, \(p<0.05\)). As shown in Fig.~\ref{fig:paper_index}b, the During\_guideline\_collab group (n=326) achieved significantly higher disruptive index than both Pre\_guideline\_collab (n=430; one-sided Brunner-Munzel test, \(p<0.001\), Cliff's Delta = 0.24) and Non\_collab (n=4,754; one-sided Brunner-Munzel test, \(p<0.01\), Cliff's Delta = 0.10). In contrast, the Pre\_guideline\_collab group had significantly lower disruptive index than the Non\_collab group (one-sided Brunner-Munzel test, \(p<0.001\), Cliff's Delta = 0.16). Collectively, these results indicate that newly formed collaborations during the guideline period yielded the highest impact in terms of both citation performance and disruptiveness.

We next assessed whether guideline development specifically enhanced publication performance by comparing papers from the treatment group (n=101) and the control group (n=101) published between 2016 and 2021 (Fig.~\ref{fig:treatment_vs_control_metrics}). Among collab papers, the treatment group (n=790) showed significantly higher citation counts than the control group (n=337; one-sided Mann-Whitney \(U\) test, \(p<0.001\), Cliff's Delta = 0.12). However, no significant citation difference emerged for non\_collab papers (treatment, n=5,104; control, n=4,488; one-sided Mann-Whitney \(U\) test, \(p=0.433\), Cliff's Delta = 0.00). By contrast, for the disruptive index, the treatment papers (n=4,754) exceeded the control papers (n=4,213) within the non\_collab type (one-sided Mann-Whitney \(U\) test, \(p<0.001\), Cliff's Delta = 0.05). No significant disruption difference emerged among collab papers (treatment, n=756; control, n=326; one-sided Mann-Whitney \(U\) test, \(p=0.063\), Cliff's Delta = 0.06). Taken together, these findings suggest that guideline development has a pronounced impact on citation outcomes for collaboration-based research. In contrast, the disruptive index was higher for non-collaborative papers--that is, for those authored by guideline researchers in collaboration with individuals outside their usual networks--indicating that research produced in these non-collab settings tended to be more innovative.

\section*{Relationship between Guideline Researchers and Cited References}

To assess the extent to which papers authored by guideline researchers contribute to the evidence base of the guideline, we examined citation types--focusing on what we term \textit{policy self-citation.} Here, policy self-citation refers to the proportion of the guideline's cited academic papers that were authored by guideline researchers. Table~\ref{tab:5_guideline_self_citation} shows that 55.2\% (101 papers) of the academic references in the guideline are policy self-citations--a rate markedly higher than the 16.1\% typically observed in the broader academic literature (see Methods section).

In addition, to investigate the \textit{distance} between the remaining academic reference of the guideline and its researchers, we constructed a co-authorship network using Scopus data up to 2021. This network included guideline researchers, their co-authors, and even their co-authors' collaborators. Furthermore, among the academic references of the guideline, 35.9\% (65 papers) included at least one author who had directly collaborated with guideline researchers (i.e., a shortest path of 1 in the co-authorship network), while 7.73\% (14 papers) included at least one author connected to guideline researchers at a network distance of 2, indicating indirect collaborative links. Only 1.7\% (3 papers) show no connection to guideline researchers within the co-authorship network. These observations highlight that guideline researchers and their immediate professional networks provide the majority of the cited evidence.

Next, we examined whether policy self-citation changed over time. The academic papers cited as reference in the guideline range from those published in 1993 to those published in 2021 (Fig.~\ref{fig:guideline_basic_info}b), the final year of the guideline's development. Among the 183 academic papers cited as the reference 101 were published after guideline development began in 2016. As summarized in Table~\ref{tab:5_self_citation}, comparing the two periods (before 2015 vs. 2016 and after) confirmed that the share of papers authored by guideline researchers--relative to all academic papers cited in the reference--increased significantly in the post-2016 period. This finding underscores the growing influence of guideline researchers in shaping policy through their active involvement in guideline development.

\begin{table}
\centering
\begin{minipage}[t]{1.\textwidth}
\centering
\caption{Classification of papers cited in the guideline by the co-authorship network distance from guideline researchers.}
\scalebox{1}{
\begin{tabular}{l c}
\hline
\textbf{Paper Classification} & \textbf{Count}  \\ \hline
Distance 0 (Policy self-citation)  & 101 (55.2\%)\\ 
Distance 1 (Shortest path = 1) & 65 (35.9\%)\\ 
Distance 2 (Shortest path = 2) & 14 (7.73\%)\\ 
Not connected & 3 (1.7\%)\\ \hline
\end{tabular}}
\label{tab:5_guideline_self_citation}
\end{minipage}%

\hfill
\begin{minipage}[t]{1.\textwidth}
\centering
\caption{Comparison of policy self-citation ratio by pre- and during the guideline development period, where the ratio represents the share of academic papers cited in the guideline that were authored by guideline researchers.}
\scalebox{1}{
\begin{tabular}{l c c}
\hline
\textbf{Period}     & \textbf{Num of Ref.} & \textbf{Policy Self-citation}  \\ \hline
Before 2015       & 82              & 40 (48.8\%)     \\ 
2016 and after       & 101             & 61 (60.4\%) \\ \hline
\end{tabular}}
\label{tab:5_self_citation}
\end{minipage}
\end{table}

\section*{Discussions}

This study examines whether involvement in an international policy guideline enhances researchers' careers, as measured by academic and policy impact, collaboration types, and policy self-citation. Our analysis of the WHO global air quality guideline reveals that researchers engaged in the guideline development process experienced a marked increase in both academic and policy citation counts compared to a matched control group. Moreover, our network analysis shows that these researchers formed new collaborations during the guideline development period, with papers emerging from these new partnerships exhibiting higher citation counts and greater disruptive index. Furthermore, and interestingly, a substantial share of the guideline's references drew on the researchers' own publications, highlighting their pivotal influence in shaping the evidence base for policymaking.

These findings provide quantitative, causally informed evidence that complements previous survey-based studies\cite{singh2019researcher, filyushkina2022engaging}. Earlier research indicated that researchers primarily engage in policy-related activities due to intrinsic motivations for societal contribution rather than institutional rewards\cite{singh2019researcher}. Additionally, while many early-career researchers are interested in being involved in policy engagement for its potential societal and professional benefits, they often face barriers such as insufficient funding, limited training, and lack of institutional support\cite{filyushkina2022engaging}. Our results extend this literature by empirically demonstrating that involvement in policy guideline development does indeed yield measurable academic and policy-related career benefits. These findings highlight the value of formally recognizing policy engagement within academic evaluation frameworks. Institutions, therefore, could consider incorporating policy contributions into researcher assessment criteria and developing structured support systems--such as training opportunities and dedicated resources--to further encourage and facilitate researchers' active involvement in science-policy interfaces.

The co-authorship network analysis further supports the notion that guideline development fosters a collaborative research environment. We observed a pronounced expansion in collaboration during the 2016--2021 period--an effect that was absent in the control group. This suggests that the process of developing international guidelines not only consolidates existing academic relationships but also promotes the formation of new partnerships. Such collaborations likely facilitate the integration of diverse expertise, thereby enhancing the overall quality and innovativeness of research outputs. This result is consistent with prior findings that fresh team configurations can generate novel ideas \cite{zeng_FreshTeamsAre_2021}. Additionally, papers resulting from collaborations among guideline researchers during the development period received significantly higher citation counts compared to those authored by similarly performing peers. One possible explanation for this mechanism is that researchers selected from various countries and regions, each bringing diverse perspectives, skills, and resources, may contribute to increased creativity and innovation in collaborative teams. However, further investigation is required to confirm these proposed mechanisms.

Furthermore, our analysis of citation types revealed that papers authored by guideline researchers constituted a substantial majority of the guideline's cited references (i.e., policy self-citation). This finding highlights the pivotal role these researchers play in selecting and synthesizing evidence during the guideline development process, potentially enhancing their influence on policy outcomes and, consequently, their career advancement. This finding indirectly supports prior empirical evidence showing that papers cited in policy documents tend to attract higher numbers of academic citations\cite{bornmann2022relevant}, suggesting a mutually reinforcing relationship between policy relevance and scholarly impact. While academic self-citation typically reflects incremental research progression and knowledge continuity\cite{aksnes_MacroStudySelfcitation_2003,hyland_SelfcitationSelfreferenceCredibility_2003,hyland_ChangingPatternsSelfcitation_2018,pislyakov_SelfcitationItsImpact_2021}, policy self-citation during guideline development may reflect practical considerations unique to policymaking contexts. Because guideline researchers are selected by organizations like WHO specifically for their established expertise, their prior research naturally becomes central to the evidence base. Thus, policy self-citation in this context should not be viewed negatively; rather, it indicates practical considerations and specialized demands of evidence-based policy formulation, distinct from academic contexts where excessive self-citation might be criticized. Nonetheless, transparency and explicit criteria for citation selection remain important to ensure diversity of evidence and to mitigate the risk of overlooking other valuable contributions. Further investigation is necessary to clarify the motivations behind this observed pattern. 

Despite these promising results, our study has several limitations. First, our analysis focuses on a science-based regulatory guideline in the domain of air quality, leaving open the question of whether similar patterns of knowledge transfer and career impact would hold in other scientific fields or for different types of guidelines. Second, as we focused exclusively on the 2021 WHO global air quality guidelines, additional analyses of other international guidelines, policy documents, or collaborative projects are necessary to verify the generalizability of our findings. Lastly, in the matching process, we did not account for other activities that researchers typically engage in--such as participation in academic conferences or involvement in broader societal contribution activities. Consequently, broadening the scope of future inquiries to include diverse science-policy interactions and a range of research disciplines is essential for a more comprehensive understanding of how involvement in policymaking influences researchers' careers.

Overall, this work has two main implications for both personal and institutional aspects. First, it provides empirical evidence that involvement in international policy formulation directly benefits researchers by enhancing their academic and policy-related impact. This quantitative demonstration adds critical support to existing survey-based observations, confirming the career benefits of researchers' active engagement in policy processes, and offers clear incentives for researchers to involve in international policymaking. Second, our findings suggest that countries and research institutions can derive strategic benefits by actively supporting and sending their affiliated researchers to participate in international guideline development, thereby not only enhancing their research impact but also contributing to the development of more innovative science. Such engagement provides researchers with platforms to demonstrate their expertise, enhance citation impact, and gain valuable collaborative opportunities. Moving forward, extending this line of research to multiple policy domains, regional contexts, and longer time horizons will deepen our understanding of the diverse ways in which research and policy can mutually reinforce each other, ultimately benefiting both scientific progress and societal decision-making.

\section*{Methods}
\phantomsection
\label{sec:methods}
\subsection*{Data}

This study focuses on the single WHO global air quality guideline published in 2021, entitled \emph{WHO global air quality guidelines: particulate matter (PM\textsubscript{2.5} and PM\textsubscript{10}), ozone, nitrogen dioxide, sulfur dioxide and carbon monoxide} \cite{world2021global}. Guideline development took place between 2016 and 2021. To identify the participants and references informing this guideline, we extracted citation information and expert names directly from the guideline text, matching them by DOI with the Scopus dataset. 

From this process, we identified 131 participants: 20 individuals from the WHO Steering Group, 26 from the Guideline Development Group, 18 from the Systematic Review Team, 2 External Methodologists, and 65 from the External Review Group. For our analysis, we refer to the 111 individuals excluding the WHO Steering Group as guideline researchers. Of these, 108 could be located in Scopus, while the remaining 3, affiliated with environmental NGOs, consultancies, or government agencies, had insufficient publication records. Among the 312 references cited in the guideline, 186 were academic papers and 126 were policy documents or books; 183 of those academic papers were successfully matched with entries in Scopus.

We then combined Scopus with Overton to build the primary dataset. Scopus provided publication and author information for 88,226,156 scholarly papers and 47,652,638 researchers spanning 1970 to 2023, while Overton contained 293,800 policy documents from intergovernmental organizations (IGOs) between 1970 and August 2024, including 6,838,694 cited articles. For each researcher, we used Scopus to calculate annual publication vectors and yearly citation counts, and we integrated Overton to determine how frequently their papers were cited in IGO policy documents.

\subsection*{Matching Similar Researchers}

We sought to test whether guideline development influences researchers' careers by pairing guideline researchers with peers who had similar research interests and performance metrics immediately prior to 2016. The matching followed three steps (Fig~\ref{fig_yearly_citation}a):
\begin{enumerate}
    \item We generated text (title or abstract) embeddings (1536-dimensional vectors) for all Scopus publicationsfrom 2011 to 2015 using \texttt{text-embedding-ada-002} and then derived specialization vector for each researcher by averaging the embeddings of their publications.
    \item For each guideline researcher, we retrieved the top 20 non-guideline researchers whose specialization vectors had the highest cosine similarity.
    \item From these 20 candidates, we calculated the mean squared error (MSE) based on three metrics: cumulative academic citations (2011--2015), cumulative policy citations (as of 2015), and cumulative publication counts (as of 2015). The candidate with the lowest MSE served as the control match.
\end{enumerate}

Of the 108 guideline researchers, 101 had publication records in Scopus between 2011 and 2015; these constituted our treatment group. We then selected 101 unique control researchers to avoid duplicate matching. A two-tailed Mann-Whitney \(U\) test at the 5\% level confirmed no significant differences between the treatment and control groups across the matching variables, indicating well-aligned research profiles (see Supplementary for details, Fig.\ref{fig:matching_metrics}).

\subsection*{Statistical Significance of Attribute Assortativity Changes (Permutation Test)}

To determine whether the observed shift in role-based assortativity--from the period before guideline development to the period after--was statistically significant, we conducted a permutation-based randomization test. First, we computed the assortativity coefficients\cite{newman2003mixing} among nodes (researchers) present in both periods and derived the observed difference. Next, we randomly permuted researcher role labels 1,000 times, recalculated the assortativity difference for each permutation, and then calculated the proportion of permuted differences that met or exceeded the observed value. The resulting \(p\)-value was adjusted using a continuity correction to avoid zero-division artifacts.

\subsection*{Definition of Collaboration}

We defined a collaboration as co-authorship among guideline researchers, treatment group or control group. Specifically, if two or more the same group researchers co-authored a paper, we classified that publication as collaborative. Otherwise, we labeled it non-collaborative. For each pair of researchers, we recorded the earliest year in which they co-authored a paper; all subsequent papers involving the same pair were assigned this same start year. We then grouped start years into three intervals: Pre\_guideline\_collab (2005--2015), During\_guideline\_collab (2016--2021), and Post\_guideline\_collab (2022--2023). For example, a 2017 publication authored by a researcher pair that first collaborated in 2011 would be labeled as \textit{Pre\_guideline\_collab}.

\subsection*{Calculation of Publication Performance Metrics}\label{sec:5_define_index}

To assess publication performance, we employed two metrics: the two-year citation count (\(C_2\)) and the disruptive index (\(D\)). Citation counts were log-transformed (\(\log(C_2 + 1)\)) to reduce skewness and approximate normality \cite{radicchi_UniversalityCitationDistributions_2008}. The disruptive index quantifies the extent to which a paper introduces novel ideas, calculated based on subsequent citations: positive values (\(D > 0\)) indicate disruptive impact, while negative values (\(D < 0\)) suggest developmental impact\cite{funk_DynamicNetworkMeasure_2017, wu_LargeTeamsDevelop_2019}. We used citation data through 2023 to compute these metrics from Scopus dataset.

\subsection*{Academic Self-Citation Ratio}
To contextualize policy self-citation rates, we calculated the average academic self-citation rate for air quality papers. Specifically, we analyzed papers in Scopus whose titles or abstracts contained any of the following terms: "air pollution," "air quality," "air pollutant," "air pollutants," "particulate matter," "PM2.5," "PM10," "ozone," "O3," "nitrogen dioxide," "NO2," "sulfur dioxide," "SO2," "carbon monoxide," or "CO." Across this corpus, the average self-citation ratio was 16.1\%. This figure serves as a benchmark against which we compare the higher rates observed within the WHO global air quality guideline itself.

\section*{Acknowledgements}
The authors thank C. Miura, K. Kimita, and all members of our research group for their invaluable comments.

\section*{Author contributions statement}
Y.T. conceived the project; Y.T. collected data and performed analysis; Y.T. wrote the manuscript, K.N., K.A., I.S. supervised the project, all authors discussed results and edited the manuscript.

\section*{Data availability}
The data that support the findings of this study are available from Elsevier but restrictions apply to the availability of these data, which were used under license for the current study, and so are not publicly available. Data are however available from the authors upon reasonable request and with permission of Elsevier.
Data requests will be handled by the corresponding author, Yuta Tomokiyo

\addtocontents{toc}{\protect\setcounter{tocdepth}{-1}}
\bibliography{references}
\addtocontents{toc}{\protect\setcounter{tocdepth}{2}}

\newpage

\section*{\huge Supplementary information}

\setcounter{figure}{0} 
\renewcommand{\thefigure}{S\arabic{figure}}
\renewcommand{\thetable}{S\arabic{table}}


\section*{Geographic and Temporal Distribution of Guideline Contributors and Citations}
\label{subsec_sup:limitation_data}

Fig.~\ref{fig:guideline_basic_info}a presents the international distribution of guideline researchers by country and region. The data reveals a pronounced geographic imbalance in expertise contribution. The United Kingdom (19) emerges as the single largest contributor, followed by the United States (14) and China (9). There is strong representation from Western Europe with the Netherlands (5), Germany (5), Argentina (5), Italy (4), and Spain (4) contributing multiple experts. Participation from Africa, South America, and South Asia is comparatively limited, with South Africa (3), Argentina (5), and India (4) serving as the main contributors from their respective regions. Additional participation was documented from Oceania, the Middle East, Central America, and Southeast Asia. This distribution demonstrates a notable geographic concentration of expertise in guideline development.

Fig.~\ref{fig:guideline_basic_info}b displays the publication year distribution of 183 the guideline-cited academic papers successfully matched to the Scopus dataset (out of 186 total citations). The orange-highlighted period (2016-2021) indicates the guideline development timeframe. Of the cited literature, 82 papers were published before 2016, while 101 papers were published during or after 2016. The papers cited in the guideline range from those published in 1993 to those published in 2021, the final year of the guideline's development.

\begin{figurehere}
    \centering
    \includegraphics[width=1.\linewidth]{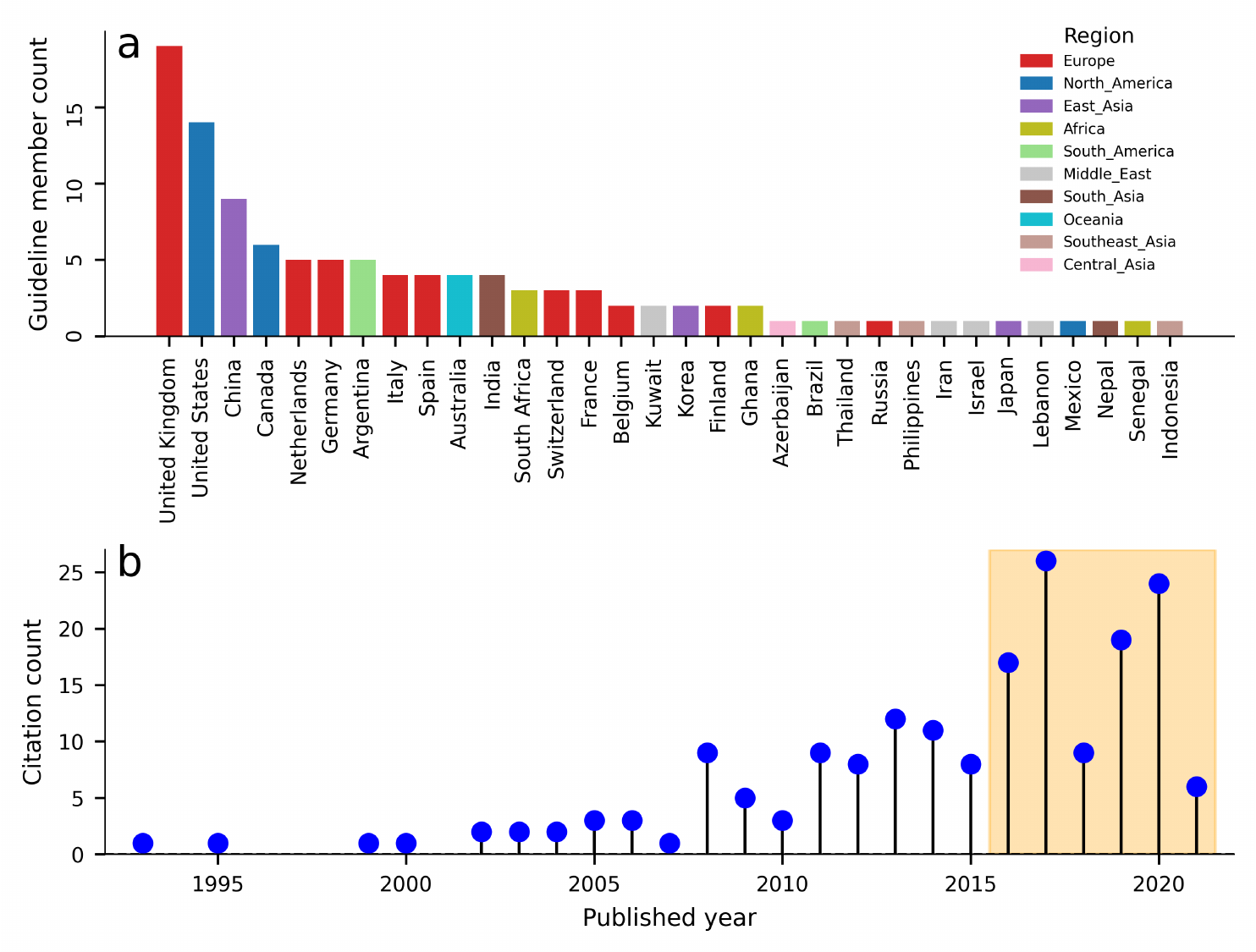}
   \caption{Geographic distribution and citation counts of the guideline reference papers. \textbf{a} Institutional affiliation by country for guideline researchers, colored by geographic region (n=101). \textbf{b} Publication year distribution of academic papers cited in the guideline (n=183), with guideline development period (2016-2021) highlighted in orange.}
   \label{fig:guideline_basic_info}
\end{figurehere}

\section*{Validation of Researcher Matching Procedure}

Our matching algorithm successfully identified control researchers with equivalent research profiles to the guideline researchers prior to guideline development. As shown in Fig.~\ref{fig:matching_metrics}, there were no significant differences between the treatment and control groups in any of the matching variables, including annual academic citation counts from 2011-2015, IGO policy citations as of 2015, and publication numbers as of 2015 (Mann-Whitney \textit{U} test, two-tailed, \(p>0.05\) for all comparisons). The close alignment of both the central tendency measures and distribution ranges confirms the validity of our control group selection, ensuring that subsequent analyses reflect the impact of guideline development rather than pre-existing differences in research performance.

\begin{figurehere}
    \centering
    \includegraphics[width=0.75\linewidth]{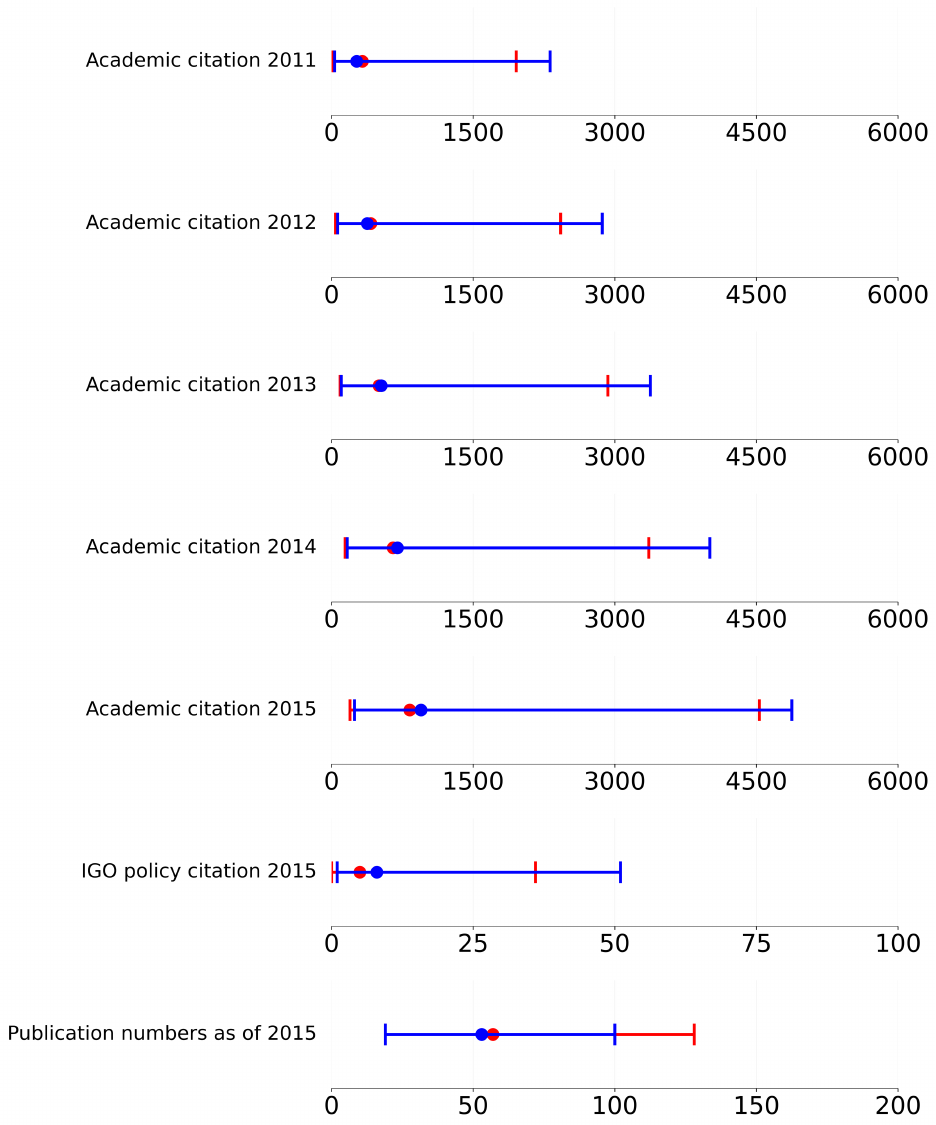}
   \caption{Pre-guideline research profiles of treatment and control groups. Annual academic citation counts from 2011 to 2015, IGO policy citations as of 2015, and the total number of publications as of 2015 are shown for treatment researchers (blue) and matched controls (red). Each point indicates the group median, with horizontal bars representing the interquartile range. No significant differences were observed across any metric (two-tailed Mann-Whitney $U$ test, $p>0.05$), confirming successful matching of control researchers prior to guideline development.}
   \label{fig:matching_metrics}
\end{figurehere}

\section*{Treatment vs.\ Control Metrics for Papers Published in 2016--2021}
\label{sec_sup:probability}

We compared publication metrics between the treatment and control groups for papers published from 2016 to 2021, using the Mann-Whitney \(U\) test. Among collaborative papers, Fig.~\ref{fig:treatment_vs_control_metrics}a shows that the treatment group (n =790) exhibited significantly higher citation counts compared to the control group (n =337) (one-sided Mann-Whitney \(U\) test, \(p<0.001\); Cliff's Delta =0.12). By contrast, no significant difference emerged for non-collaborative papers(treatment group: n=5,104; control group: n=4,488) (one-sided Mann-Whitney \(U\) test, \(p=0.433\); Cliff's Delta =0.00).

In terms of the disruptive index (Fig.~\ref{fig:treatment_vs_control_metrics}b), non-collaborative papers in the treatment group (n =4,754) attained significantly higher values than those in the control group (n =4,213) (one-sided Mann-Whitney \(U\) test, \(p<0.001\); Cliff's Delta =0.05). However, collaborative papers revealed no significant difference in disruptive index between the treatment (n =756) and control (n =326) groups (one-sided Mann-Whitney \(U\) test, \(p=0.063\); Cliff's Delta =0.06).

Taken together, these findings suggest that participation in guideline development most notably enhances the citation impact of collaborative publications, whereas its effect on disruptive index appears more limited.

\begin{figurehere}
    \centering
    \includegraphics[width=1.0\linewidth]{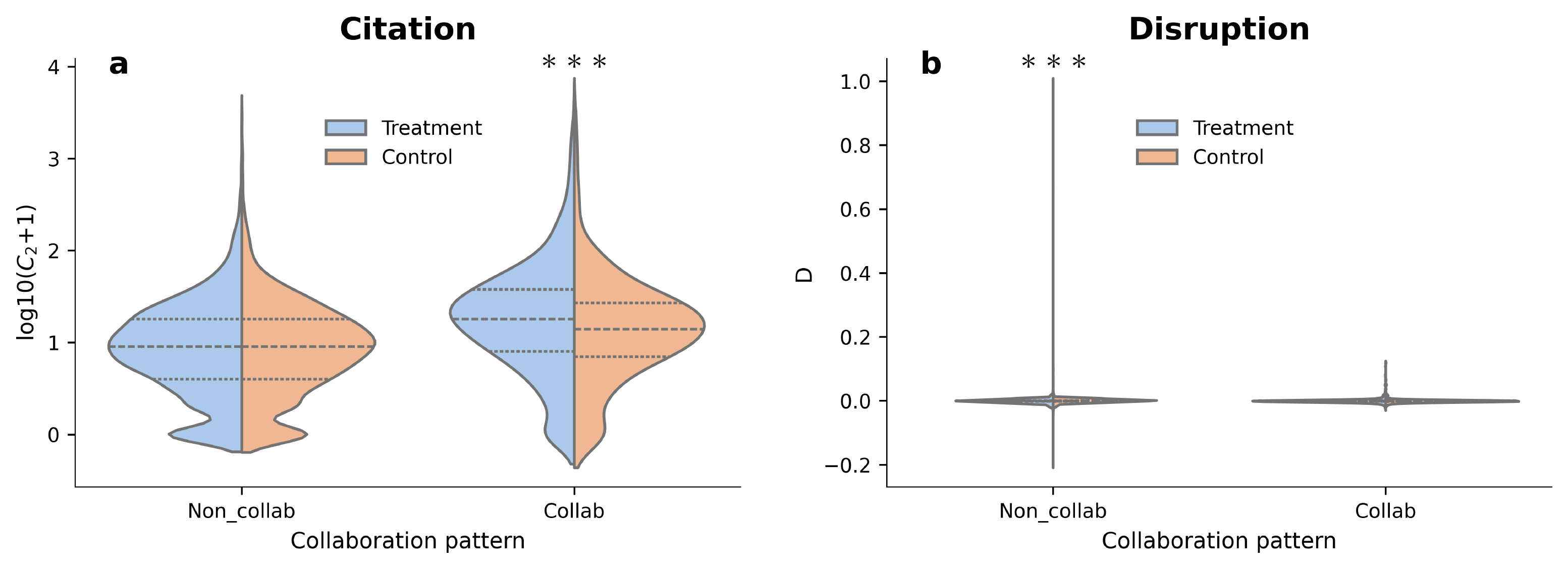}
   \caption{Impact metrics for treatment vs.\ control papers published during 2016--2021.Violin plots depict \textbf{a}~log-transformed two-year citation counts, \(\log_{10}(C_2 + 1)\), and 
\textbf{b}~disruptive index \((D)\) for non-collaborative (\textit{Non\_collab}) and collaborative (\textit{Collab}) papers. 
The treatment group results are shown in blue, and the control group in orange. Horizontal dashed lines represent the quartiles. }
   \label{fig:treatment_vs_control_metrics}
\end{figurehere}
\end{document}